\documentclass{stsci_report}

\usepackage{graphicx}
\usepackage{siunitx}
\usepackage{todonotes}
\usepackage{pdfpages}
\usepackage{booktabs}
\usepackage{caption}
\usepackage[hidelinks]{hyperref}
\usepackage{amssymb}

\copyrighttext{Copyright\copyright\ \the\year\ The Association of Universities for Research in Astronomy, Inc. All Rights Reserved.}

\presubtitle{Instrument Science Report WFC3 2025-06}

\title{Updates to the WFC3/UVIS Saturation Map}

\author{Mitchell Revalski, Isabel Rivera, Varun Bajaj, Frederick Dauphin}

\date{August 20, 2025}

\begin{document}

\maketitle

\abstract{The \texttt{calwf3} software for WFC3/UVIS utilizes a reference file to flag pixels that are saturated beyond their full-well depth. Previously, this was accomplished using a constant threshold of 65,500~e$^{-}$ across the entire detector. In this study, we retrieved $\sim$1~million stars from the Mikulski Archive for Space Telescopes (MAST) to determine the flux level at which the Point Spread Function begins to flatten, which occurs as the central pixel saturates. We quantified the saturation limit as a function of position on the detector in 1,024 discrete regions, and interpolated to a pixel-by-pixel saturation map to construct a spatially-variable saturation map reference file that is now implemented in the \texttt{calwf3} calibration pipeline. We find the saturation varies by 13\% across the UVIS detectors, from 63,465~e$^{-}$ to 72,356~e$^{-}$. These values agree well with earlier studies using sparser datasets, with the current analysis leading to improved characterization on small scales. Critically, the revised saturation values are larger than the previous constant threshold over 87\% of the UVIS detector, leading to the recovery of usable science pixels near bright sources. This update greatly improves the robustness of saturation flags in the Data Quality arrays of observations obtained with WFC3/UVIS, and users are encouraged to redownload their data from MAST to benefit from the improved flags.}

\section*{Introduction}

The Wide Field Camera 3 (WFC3) onboard the Hubble Space Telescope (HST) has captured revolutionary images of the cosmos since its installation in 2009. As a Charge-Coupled Device (CCD), WFC3's Ultraviolet and Visible (UVIS) channel uses a silicon detector that converts incident photons into electrons, which are confined in the potential well of each pixel by an applied voltage. While CCDs are advantageous for their high quantum efficiency, large dynamic range, and linear response to light, there is also a physical limit to the number of electrons that each pixel can hold before they overwhelm the potential barrier and spill into adjacent pixels. This limit is known as the full well depth, and pixels exceeding this limit are considered saturated. When saturation occurs, charge spills into adjacent pixels, and the pixel's response to light is no longer linear with exposure time. It is therefore critical to characterize the full well depth across the detector to understand when the photometry, astrometry, and morphology measurements of astronomical sources may be compromised.

Initially, a conservative saturation limit of 65,500~e$^{-}$ was chosen, which is approximately equal to the limit of a 16-bit number at a gain of unity (65,535). Ground-based testing before launch determined that pixels saturated at $\sim$70,000 e$^-$ \parencite{Bushouse2006}, with spatial variations across the detector caused by variations in the thickness of the silicon substrate as shown in Chapter 5.6, Figure 5.19 of the WFC3 Data Handbook \parencite{Pagul2024}. These variations were characterized by \textcite{Gilliland2010}, where they compared the fluxes of stars before and after saturation using short and long exposures of the same fields obtained without dithering. By measuring the fluxes of stars in unsaturated exposures and comparing them to their corresponding saturated measurements, they established a relationship between exposure time and the number of saturated pixels to determine the value at which saturation first occurs. This allowed \textcite{Gilliland2010} to characterize spatial variations in the saturation using 313 stars across the detector, which they interpolated and smoothed to a pixel-by-pixel map using a large 400~$\times$~400 pixel Gaussian kernel. While this method is robust, it requires undithered short and long exposures, tedious defining of photometric apertures, and a reliable linearity correction for fluxes beyond saturation, which can be significant for portions of the UVIS1 chip (see \cite{Gilliland2010} for details). At that time, the resources were not available to make the extensive modifications required for \texttt{calwf3} to use a spatially-dependent map, which was later completed by \textcite{Rivera2023}.

In revisiting this analysis in 2025, we are fortunate to have many years of science and calibration observations, with an easily accessible catalog of well-characterized stars available through the Mikulski Archive for Space Telescopes (MAST) stellar cutout database described in \textcite{Dauphin2021}\footnote{\url{https://www.stsci.edu/hst/instrumentation/wfc3/data-analysis/psf/psf-search}}. This MAST collection enables users to retrieve stellar cutouts for tens of millions of stars observed with WFC3. Queries can easily be sorted by telescope instrument, filter, date, and the stellar parameters measured using the HST1PASS software \parencite{Anderson2022}, including peak and total flux, quality of fit, focus model value, and location on the detector. In addition, users can retrieve these stars from data products at different calibration levels including uncalibrated RAW files, calibrated, flat-fielded FLT files, and Charge Transfer Efficiency (CTE) corrected FLC files. In this study, we leverage the MAST cutout database to robustly quantify the saturation threshold across the WFC3/UVIS detector. Specifically, we use stars from the database to precisely determine the flux level where the shape of the Point Spread Function (PSF) breaks at each location on the detector, corresponding to the flux at which the central pixel first begins to saturate.

\section*{Methods}

Our methodology is fundamentally based on characterizing how the PSF changes for stars before and after the central pixel saturates. In simple terms, the PSF describes how the light from an unresolved source is spread across multiple detector pixels after traversing the telescope's optical light path. The largest amount of flux will be contained in the pixel where the star is located, with less flux contained in surrounding pixels, decreasing with distance from the central pixel. The shape of this light distribution characterizes the system's PSF. As the detector accumulates star light, the flux in the central pixel and those around it will increase in a distribution matching the PSF. As the central pixel saturates, the rate at which it accumulates flux will slow dramatically, flux will begin to spill into adjacent pixels, and the light distribution will no longer match the shape of the intrinsic (unsaturated) PSF.

\begin{figure}[b!]
\vspace{-0.5em}
\centering
\includegraphics[width=\textwidth]{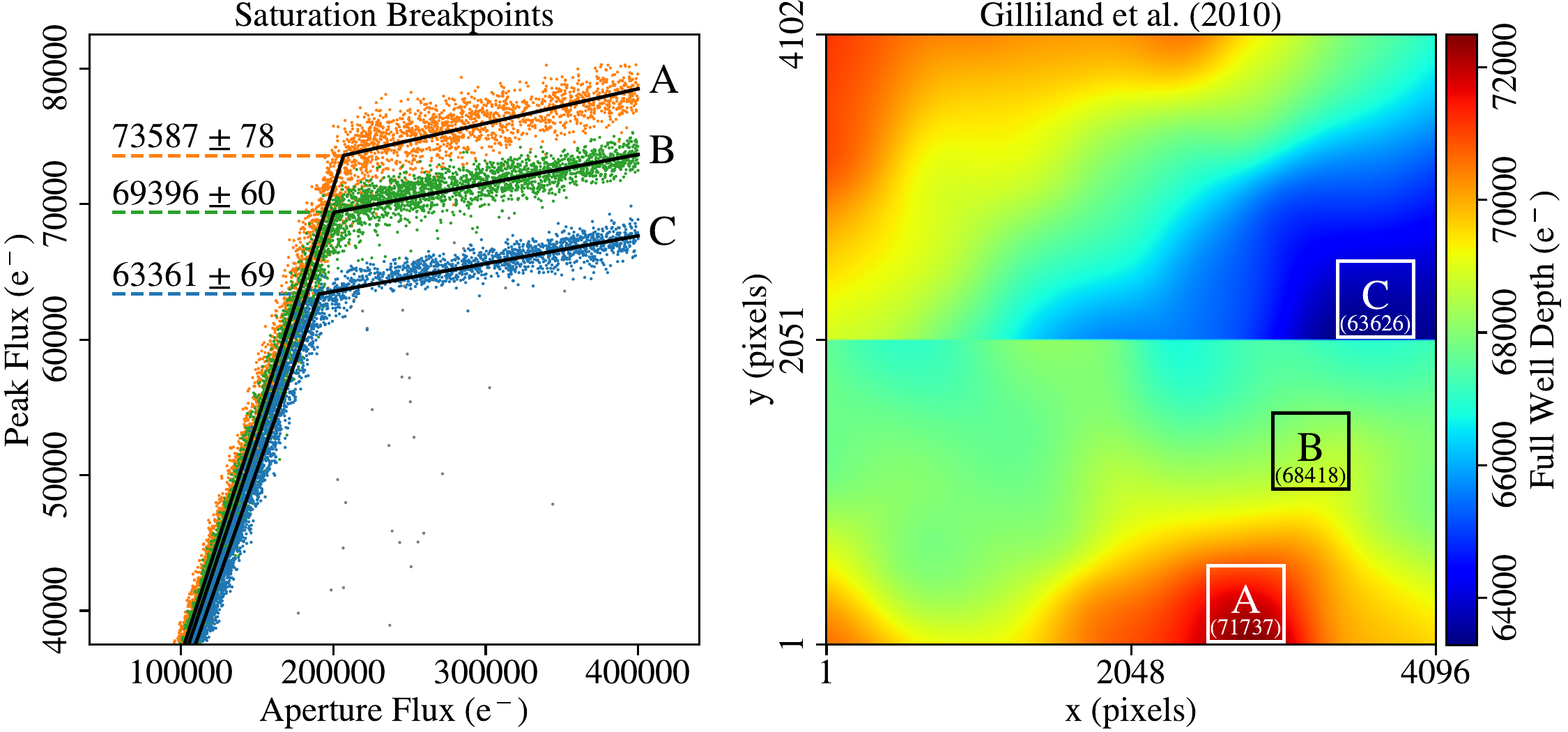}
\vspace{-1.5em}
\caption{A demonstration of our procedure for determining how the saturation limit varies across the WFC3/UVIS detector. In the left panel, we show the peak flux (ordinate) versus 3$\times$3 aperture flux (abscissa) for thousands of stars retrieved from MAST, which were observed at three discrete locations on the detector (A, B, C). As the central pixel saturates, the rate at which it accumulates flux slows dramatically, leading to a breakpoint that we fit using a piecewise linear function, with sigma-clipped outliers shown by gray points. In the right panel, we show the 2D saturation map determined by \textcite{Gilliland2010} using a related technique. The regions labeled A, B, and C correspond to the 512$\times$512 pixel detector regions for the stars that are fit in the left panel. The best-fit breakpoints are very similar to the medians of each corresponding region in the \textcite{Gilliland2010} map. This serves as a proof of concept, and the small differences are expected as discussed in the Appendix.}
\label{fig:methods}
\vspace{-1em}
\end{figure}

As shown in Figure~\ref{fig:methods}, we can quantify this breakpoint very simply without a PSF model by comparing the fraction of flux in the central pixel to an aperture of 3$\times$3 pixels around it. This ratio is constant as flux accumulates until the central pixel reaches saturation, at which point the ratio changes sharply. In this study, we use this breakpoint to characterize the saturation level at regular intervals across the WFC3/UVIS detector. This is similar to the method used by \textcite{Cohen2020} to measure variations across the ACS/WFC.

We perform our analysis on identical stars in the RAW and CTE-corrected FLC files. The RAW analysis is required to develop the \verb|calwf3| calibration file, as saturation map flagging is performed before most other calibrations (see Figure 2 of \textcite{Rivera2023} for a diagram of the pipeline steps). RAW files are in units of Data Number (DN) and so differ in value from the FLC files by the gain plus additional factors like the flat-fielding. We also perform the analysis on the FLC files, as these are calibrated images in units of electrons that provide a familiar format for comparison with \textcite{Gilliland2010}, and other instruments such as ACS/WFC that has overall higher saturation limits \parencite{Cohen2020}.

First, we query the MAST cutout service\footnote{\url{https://github.com/spacetelescope/hst_notebooks/tree/main/notebooks/WFC3/mast_api_psf}} for all stars matching a defined set of search criteria. We select the F814W filter as it has been used more than any other filter on WFC3/UVIS. In addition, stars are generally brighter than when using the shorter wavelength wide-band filters, and we have a focus-diverse effective PSF model \parencite{Anderson2018} for precise fitting with HST1PASS \parencite{Anderson2022}. Selection parameters related to fluxes are in native units, with identical stars selected in the FLC and RAW datasets. We adopted the following search criteria: 1) we select stars with high-quality fits (\textsc{qfit} $\leq$ 0.06) in exposures with 2) at least 10 seconds of exposure time to avoid shutter effects, and 3) a minimum central flux of 30,000 (\textsc{pixc} $\geq$ 30,000) that 4) do not have a brighter pixel within 10 pixels of the star (\textsc{hmin} $\geq$ 10) with 5) a maximum sky flux of 1000 (\textsc{sky} $\leq$ 1000) to avoid high backgrounds that may bias measurements, with 6) a maximum of 9 saturated pixels as defined using the current scheme. Stars beyond this limit are not useful for determining the saturation breakpoint, as most of the pixels around the center pixel will also be saturated.

\begin{figure}[b!]
\vspace{-1em}
\centering
\includegraphics[width=\textwidth, trim={1.2em, 0em, 1.2em, 0em}, clip]{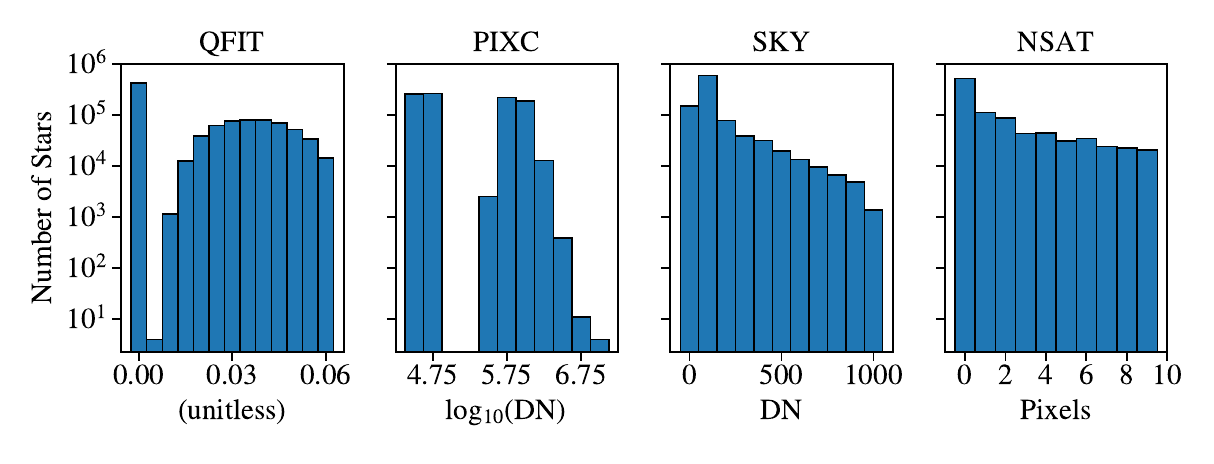}
\vspace{-2.5em}
\caption{Properties of the 924,667 stars used in the saturation map analysis. The histograms show, from left to right, the distributions for quality of fit (\textsc{qfit}), central pixel flux (\textsc{pixc}), background sky flux (\textsc{sky}), and the number of saturated pixels (\textsc{nsat}). These are displayed for stars in the RAW images and so fluxes are in Data Numbers (DN). The gap in \textsc{pixc} is a feature of how HST1PASS flags saturated stars, and which is not utilized in this analysis.}
\label{fig:histograms}
\end{figure}

We present distributions of the stellar parameters for the selected stars in Figure~\ref{fig:histograms}. The spike at low \textsc{qfit} represents the saturated stars, which are defined by HST1PASS to have \textsc{qfit} = 0. In addition, the small gap in \textsc{pixc} is due to the way HST1PASS separates unsaturated and saturated stars. Specifically, if a star is flagged as saturated, then HST1PASS attempts to determine the amount of flux that has bled into adjacent pixels and adds that flux back into the \textsc{pixc} value, leading to a natural division in the central pixel flux distribution. We also start with a minimum \textsc{pixc} $\geq$ 30,000 to balance the number of points below and above the breakpoint for equal weight fitting. While we initially set a smaller limit on the \textsc{sky} flux, this occasionally rejected stars with only a few saturated pixels due to the background aperture region. As these stars are critical for probing the turnover, we allow for a large range in \textsc{sky} flux and use an iterative sigma-clipping procedure to reject spurious sources with high \textsc{sky} fluxes due to contaminating stars, charge bleed from other sources, and other artifacts. Finally, the number of saturated pixels is defined as the number of contiguous pixels with fluxes of $>$~65,500~e$^{-}$. We only use this parameter to exclude very heavily saturated stars, as they are not useful for probing the turnover point during fitting.

\begin{figure}[b!]
\centering
\vspace{-0.5em}
\includegraphics[width=\textwidth, trim={2em, 4em, 5.2em, 9em}, clip]{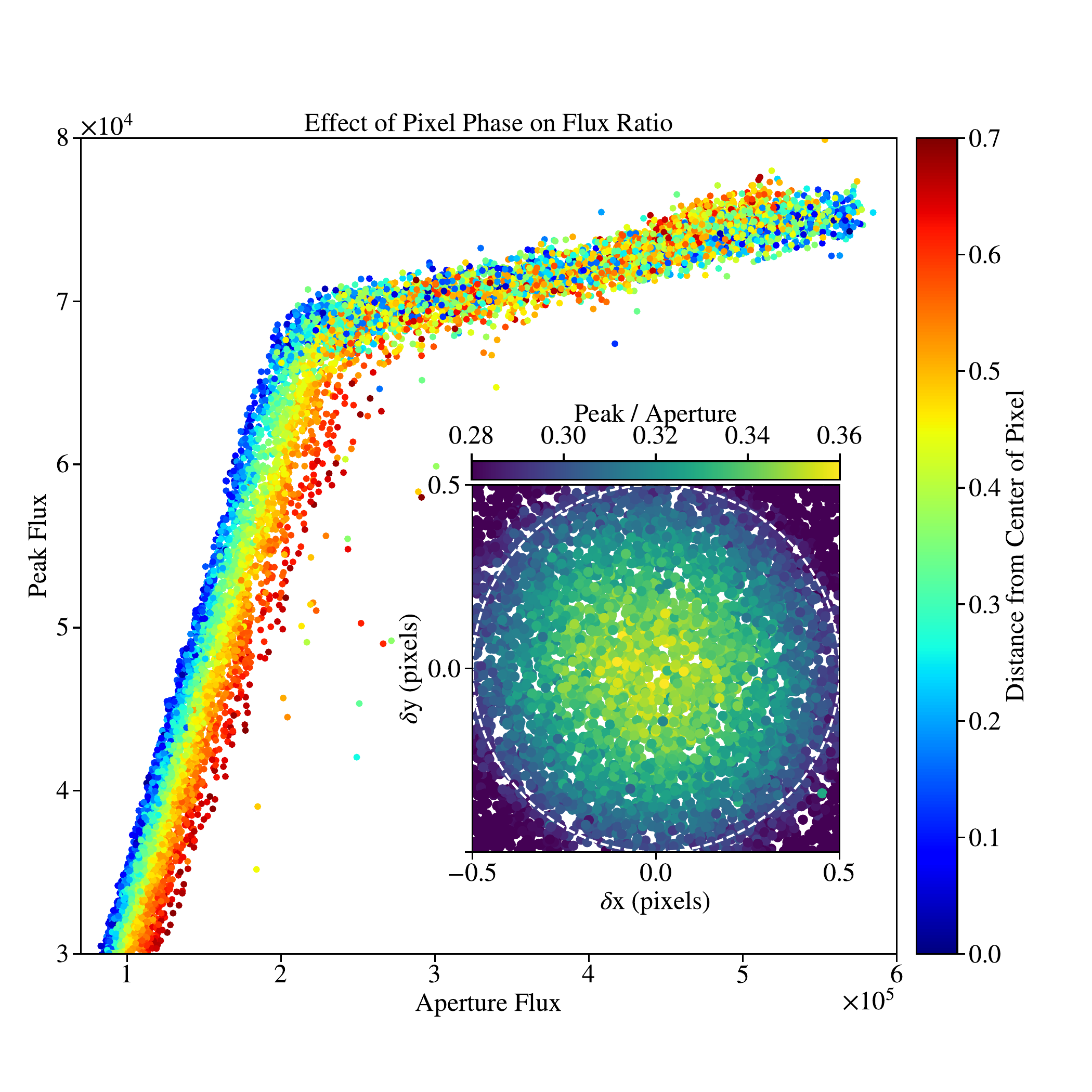}
\vspace{-2.5em}
\caption{The effect of pixel phase on the peak to aperture flux ratio. Stars landing near the center of a pixel have the highest ratio, with most of the flux contained within the central pixel. Stars landing near the edges and corners lose more flux to the surrounding pixels. The main panel shows this effect as a function of flux color-coded by pixel phase, while the inset shows the effect as a function of pixel phase color-coded by flux ratio. These panels use stars drawn from a 400$\times$400 box at (x, y) = (1024, 512) on UVIS1. Stars at $r >$~0.5 pixels (white dashed circle) bias the best-fit turnover by $>$~100 e$^{-}$ and are excluded from the analysis.}
\label{fig:phase}
\vspace{-1em}
\end{figure}

This query returns $\sim$1.24 million sources, some of which we filter out for our analysis. Specifically, we focus on stars that land near the center of each pixel (low pixel phase error) because stars near the edges reduce the dynamic range in flux between the center pixel and the surrounding 3$\times$3 aperture. As stars approach the edge of their pixel, more flux is also shifted outside of the surrounding aperture, changing the ratio between the center pixel and square aperture. As shown in Figure~\ref{fig:phase}, pixel phase has a clear effect on this flux ratio. Interestingly, by fitting stars near the extreme center and edges of pixels independently, we found that the bias is $<$~100~e$^{-}$ until stars at $r >$~0.5 pixels are included (i.e. the corners). Effectively, the turnover shifts to the right in Figure~\ref{fig:phase}, and only slightly downward. Based on these tests, we only include stars located within $r \leq $~0.5 pixels of the center of each pixel.

With the resulting sample of 924,667 stars, we optimized the spatial resolution of our map by selecting the smallest even division of the detector size that retained a sufficient number of stars for robust fitting. We found that $\gtrsim$~250 stars are sufficient for the piecewise fit to converge on a stable breakpoint. Splitting the detector into 128$\times$128 pixel regions results in $\sim$400~--~2000 stars in each box, with higher concentrations of stars observed over time appearing near the center of the detector. Dividing the detector into 128$\times$128 pixel regions yields a 32$\times$32 element grid with 1,024 unique saturation values across the detector. 

Next, we fit a piecewise linear function to the fluxes for stars in each region to identify the breakpoint where the central pixel saturates, yielding the full well depth for each location on the detector. We do this by defining a piecewise linear function that takes the 3$\times$3 aperture and peak center fluxes as x and y coordinates and conducts a sigma-clipped fitting (see Figure~\ref{fig:methods}) using standard NumPy \parencite{numpy} and SciPy \parencite{scipy} functions\footnote{Programmatically, \texttt{numpy.piecewise(x, [x < x0], [lambda x:m1*x + y0-m1*x0, lambda x:m2*x + y0-m2*x0])}, where \texttt{m1} and \texttt{m2} are the slopes. This function is then fit using \texttt{scipy.optimize.curve\_fit()}.}. Specifically, the function will produce an initial fit, calculate the standard deviations of points for each half of the piecewise fit, and reject points that are more than 5$\sigma$ away from the lines. The code repeats the fit on the outlier-rejected data, iterating until no more points are rejected or a maximum of 5 iterations is reached, whichever comes first.

We initialize the four model fit parameters to converge on a solution. These parameters are the aperture flux (x) and peak flux (y) of the breakpoint, and the slopes of the two lines. We provide an initial guess that the break is near (x, y) = (2.5$\times$10$^{5}$, 6.5$\times$10$^{3}$) e$^{-}$ for the FLCs, and divided by the gain of 1.56 for RAWs, with unsaturated and saturated slopes of 0.27 and 0.02, respectively. We found the slopes are very similar across different regions of the detector, and these starting values worked well at all locations. Finally, we save the best-fit saturation values and reconstruct them on a pixel grid matching the detector layout.

We verified that our results do not depend on the choice of filter when fitting the RAW data by generating a coarser spatial map using the smaller number of stars observed with the F336W filter. The mean and median values were identical to within $\sim$50~e$^{-}$ with differences of $<$~200~e$^{-}$ across all regions of the detector. For simplicity, and because the F814W results already capture detector-scale variations in the CCD thickness (Figure 5.19 of the WFC3 Data Handbook; \cite{Pagul2024}), we opted not to include additional filters. The saturation levels in the derived maps vary by $\sim$1000~e$^{-}$ when fitting stars observed with different filters in the FLCs, as each filter has a different flat-field applied during calibration. For these reasons, we utilize a map constructed from the RAW data for use with \texttt{calwf3}.

The most significant uncertainty that we isolated is due to time-variability. We generated coarse spatial maps using the oldest and most recent 25\% of stars in the database and found that the median of the saturation map decreases by approximately 600 DN in the RAW data and 795 e$^{-}$ for the FLC data over 15 years. The ratio of the decrease seen in the RAW and FLC is less than the UVIS gain. The result is not a strong function of distance from the detector readout between the RAW and FLC data and so is likely not attributed to CTE losses. This difference may be caused by a complex combination of detector degradation, and/or changes in the sensitivity and gain. Considering that the difference between the saturation map at any given time and the average map across all time periods is $<$~400 e$^{-}$, or $\approx$~4\% of the overall variation in saturation values across the full detector, we defer a detailed characterization of time variability to a future investigation.

\section*{Results}

\begin{figure}[b!]
\centering
\includegraphics[width=0.49\textwidth, trim={4.75em, 2em, 4.85em, 2em}, clip]{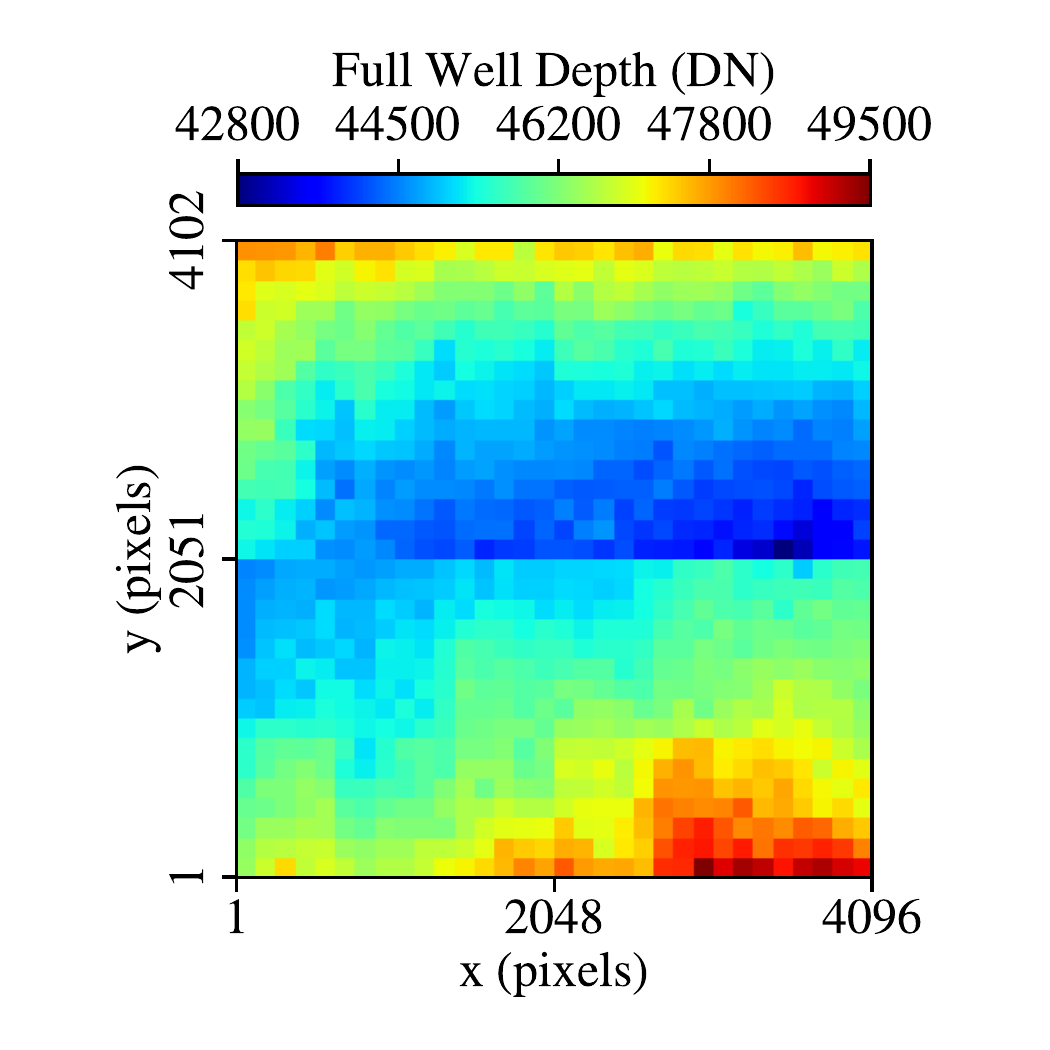}
\includegraphics[width=0.49\textwidth, trim={4.75em, 2em, 4.85em, 2em}, clip]{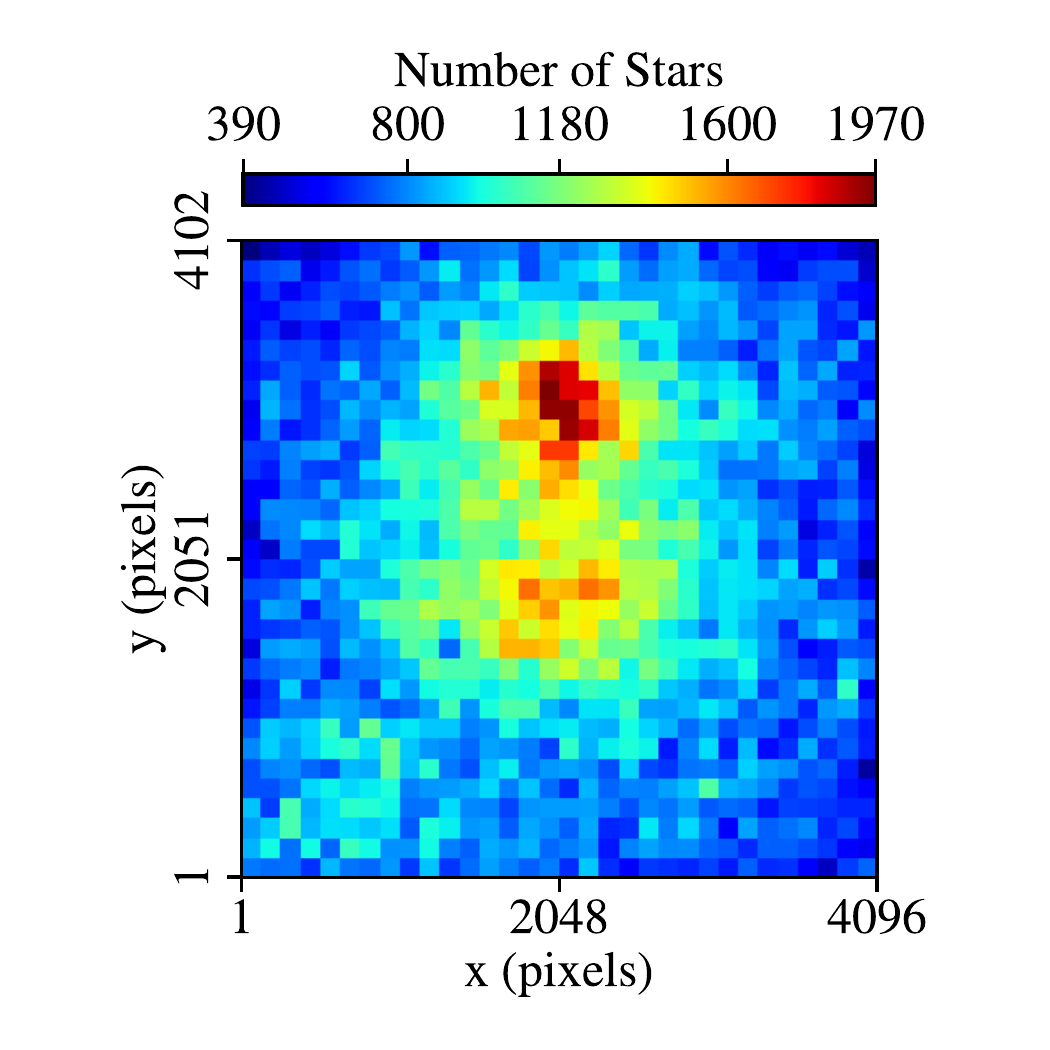}
\caption{The results of our fitting procedure with the saturation map shown on the left, and a map for the number of stars used in each fit on the right. Fitting was performed on the RAW images and so are in units of Data Numbers (DN). The high density of stars near the center of the detector is due to preferentially centering targets on each chip over 15 years of observations, while the subtle enhancement in the lower-left is due to the use of subarrays near the detector readout. There are $>$~250 stars at all locations as required for robust fits.}
\label{fig:results}
\end{figure}

We present the results of our fitting procedure in Figure~\ref{fig:results}. In the left panel, we display the saturation map that results from fitting the RAW data files in units of Data Numbers. On the right, we show a density map for the number of stars used in each piecewise linear fit. Overall, we see a similar pattern of variations to the \textcite{Gilliland2010} map shown in the right panel of Figure~\ref{fig:methods}, with subtle differences in structure due to the fact that we have fit the RAW files that have not been subject to flat-fielding and additional calibration steps. The overall large-scale structure is driven by thickness variations in the CCD substrate.

\begin{figure}[h!]
\vspace{-1.5em}
\centering
\includegraphics[width=0.99\textwidth, trim={10.7em 32em 8.5em 30em}, clip]{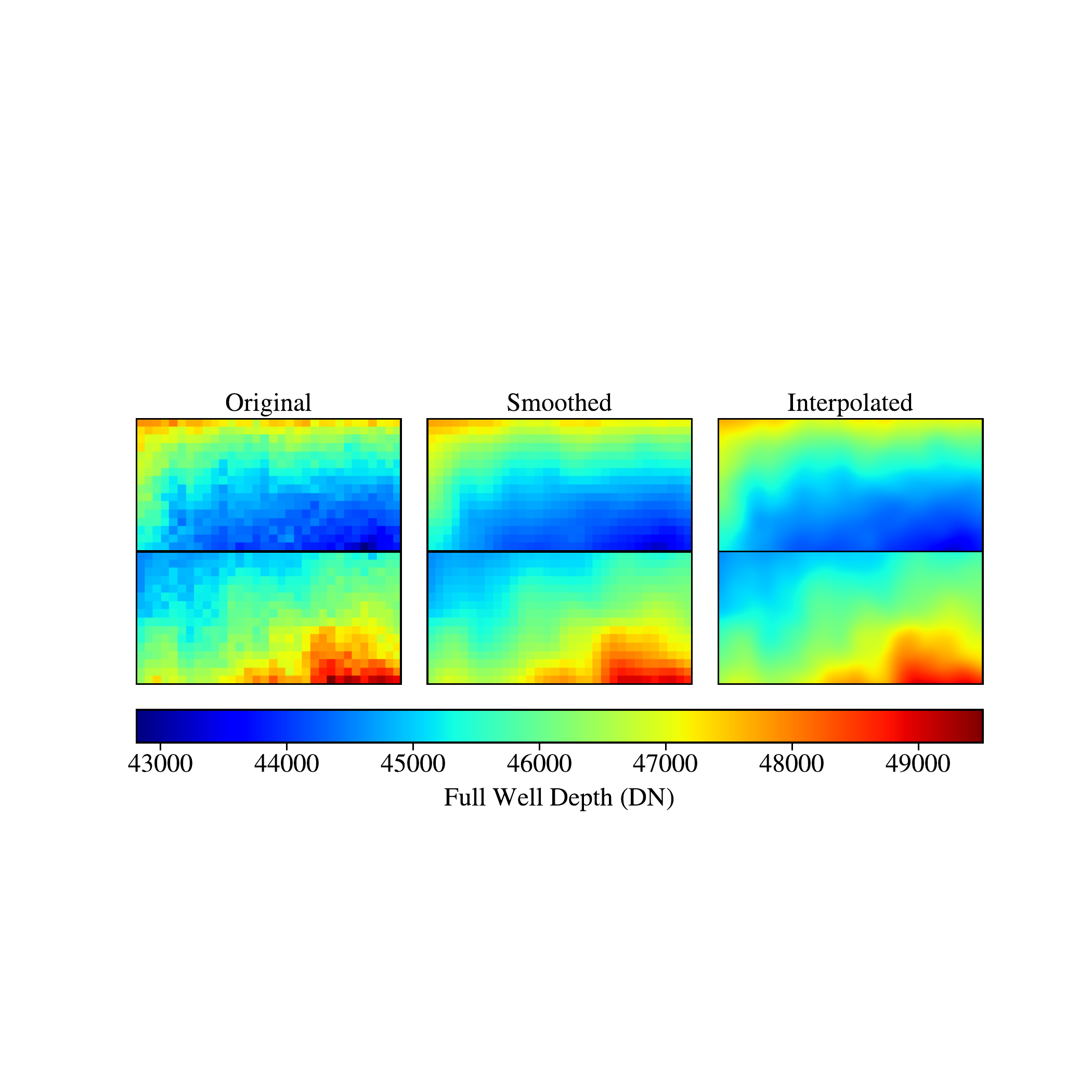}
\caption{The post-processing steps applied to the saturation map from Figure~\ref{fig:results}. We start with the original map (left panel) and apply a minimal Gaussian smoothing with a FWHM of 2 pixels (center panel) to ensure monotonic changes across adjacent boxes while still preserving large-scale variations. This ensures that interpolation to the native WFC3/UVIS plate scale on a pixel-by-pixel basis (right panel) is free of small-scale interpolation artifacts.}
\label{fig:interpolate}
\vspace{-1em}
\end{figure}

As shown in Figure~\ref{fig:interpolate}, we smooth and interpolate the saturation map to the native WFC3/UVIS grid of 4096 $\times$ 2051 pixels for each UVIS chip. Specifically, we apply a minimal smoothing using a Gaussian filter with a full-width at half maximum (FWHM) of two pixels. This yields more monotonic variations across adjacent boxes and eliminates spurious features during interpolation. We use the Scipy \parencite{scipy} \texttt{gaussian\_filter} and \texttt{RegularGridInterpolator} for this process, with a cubic order to interpolate the 128$\times$128 pixel regions to the native WFC3/UVIS pixel grid. We confirmed that the smoothing and interpolation steps do not change the minimum, maximum, or median of the distribution for each chip by more than tens of electrons.

\section*{CRDS Reference File}

The fitting procedure described in the previous section produces a map that requires further processing for compatibility with the WFC3/UVIS calibration pipeline, known as {\texttt{calwf3}\footnote{\url{https://github.com/spacetelescope/hstcal}}. Specifically, \verb|calwf3| utilizes reference files from the Calibration Reference Data System (CRDS) database. We perform the following post-processing for CRDS format compliance.

First, the saturation flagging step in the pipeline occurs after overscan (\texttt{BLEVCORR}) and residual bias (\texttt{BIASCORR}) subtraction (see Figure 2 of \textcite{Rivera2023} for a chart of the pipeline steps in \verb|calwf3| $\geq$ v3.7.1). To account for this, we subtract the commanded bias level from each amplifier quadrant in the saturation map to closely approximate the state of the data being processed at the time the flagging is applied. We note that during the calibration process, the best-fit bias level is subtracted from each amplifier quadrant of the science data, which can differ from the commanded bias levels by a few counts. In addition, a superbias file is then used to correct for residual pixel-to-pixel bias variations, which are typically less than one count, except for a few dozen pixels where the correction is tens of counts (see WFC3 ISR 2023-03). For simplicity, only the commanded bias is subtracted from the saturation map in the CRDS reference file, technically leading to minuscule offsets in the bias levels between the science arrays and saturation map. However, the magnitude of this offset is at most a few counts and has a negligible effect on the saturation flagging.

Although the map is originally derived in units of DN, the CRDS deliverable is converted to e$^-$ by applying an average commanded gain of 1.56 e$^-$/DN. This ensures consistency with the gain-corrected format for reference files at this stage in the pipeline. Internally, \verb|calwf3| converts the reference map to DN using this value before comparing them to the science data, which are in DN units at the saturation flagging step. Overscan regions are included in the map and are explicitly set to zero. Figure \ref{fig:crds_sat_map} shows the map that is now used in \verb|calwf3|.

In addition to the 1$\times$1 full-frame map, we generated 2$\times$2 and 3$\times$3 pixel binned versions by applying summation binning. While this ensures compatibility with binned science images, it is not an ideal approach for defining saturation. When multiple detector pixels are combined into a single binned pixel, it is not possible to determine whether one or more individual pixels within the bin have saturated—only the summed signal is available. As a result, determining whether a binned pixel should be flagged as saturated is ambiguous: the summed signal can exceed the saturation threshold even if none of the contributing pixels are individually saturated, or remain below it despite one or more pixels having actually saturated. Building a statistically robust saturation map directly from binned cutouts would require a more complex analysis and significantly larger sample sizes. However, because binned WFC3/UVIS observations are very rare (2$\times$2 and 3$\times$3 binning account for 0.7\% and 0.2\%, respectively, of all UVIS external observations), we do not pursue that analysis.

Programmatically, \texttt{calwf3} selects the appropriate saturation reference file based on the \texttt{DETECTOR}, \texttt{BINAXIS1}, and \texttt{BINAXIS2} header keywords in the input file's primary header. All images to which this map applies will have \texttt{DETECTOR = UVIS}, and the binning keywords instruct \texttt{calwf3} to use the 1$\times$1, 2$\times$2, or 3$\times$3 map. For subarray images, the 1$\times$1 saturation map is cropped within \texttt{calwf3} to match the subarray's size and location on the full detector.

\begin{figure}[b!]
\vspace{-1.25em}
\centering
\includegraphics[width=\textwidth, trim={1.25em, 4.75em, 3em, 6em}, clip]{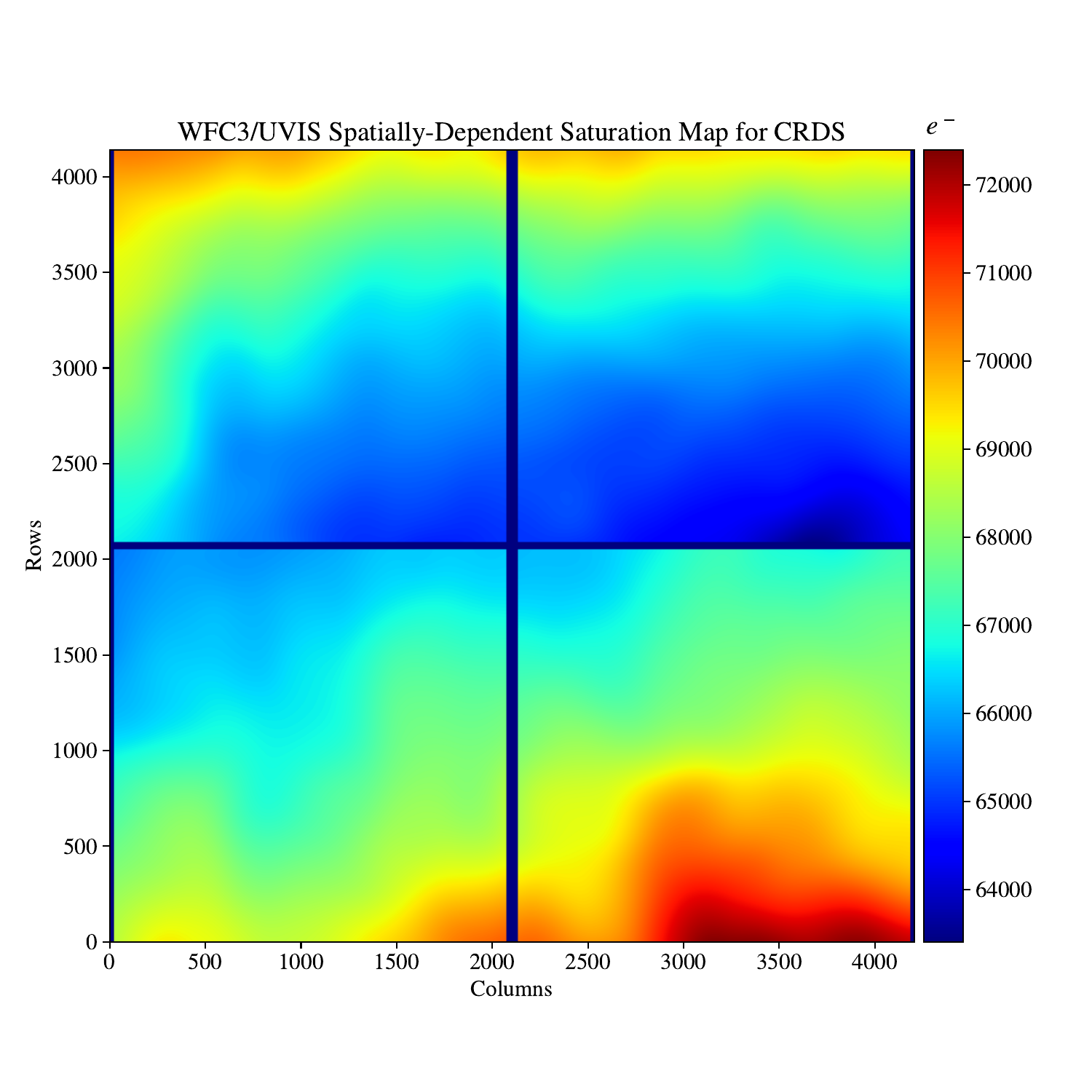}
\vspace{-1.75em}
\caption{The final CRDS-compliant saturation map now used with \texttt{calwf3}, in units of electrons. This version of the map has been bias-subtracted per quadrant, converted to electrons by applying the commanded gain (1.56), and includes overscan regions explicitly set to zero. The map reflects the detector state after overscan subtraction, effectively matching the stage at which saturation flagging is performed within the \texttt{calwf3} data calibration pipeline.}
\label{fig:crds_sat_map}
\end{figure}

This new reference file is shown in Figure~\ref{fig:crds_sat_map} and has been used to reprocess all WFC3/UVIS data. The simplest way users can obtain updated FLT, FLC, or drizzled products is by downloading them from MAST starting September 2025. Alternatively, existing RAW data can be reprocessed using the new saturation map. To do this manually, set the \texttt{SATUFILE} keyword in the file's primary header to the new reference file and run the data through \texttt{calwf3} v3.7.1 or later. Earlier versions of \texttt{calwf3} will continue to use the older, single-value threshold. If the \texttt{SATUFILE} keyword is missing or invalid, then all versions of \texttt{calwf3} will default to the single-threshold approach. For reprocessing data calibrated prior to the release of \texttt{calwf3} v3.7.1 (December 2023), see \textcite{Rivera2023}.

\section*{Validation with calwf3}

Although the capability to apply a two-dimensional saturation map was introduced in \texttt{calwf3} v3.7.1, the delivered CRDS reference file effectively continued to perform single-value flagging across the detector by design (see \textcite{Rivera2023} for details). This approach does not account for pixel-level or large-scale spatial variations in the full-well depth and can result in either overflagging or underflagging in certain regions of the UVIS detector.

To assess differences between flagging performed using the uniform and spatially variable maps, we computed a difference image by subtracting the results of the spatial map from the uniform map. This reveals regions where the pixel-level saturation thresholds derived from the spatial method are either higher or lower than the fixed value used previously. As shown in Figure \ref{fig:uni_vs_spat}, 87\% of pixels have higher thresholds in the spatial map, while 13\% have lower thresholds that are almost entirely confined to UVIS1. Overall, the spatial map recovers a large fraction of usable science pixels compared to the constant threshold map.

\begin{figure}[htb!]
\vspace{-1em}
    \centering
    \includegraphics[width=\textwidth]{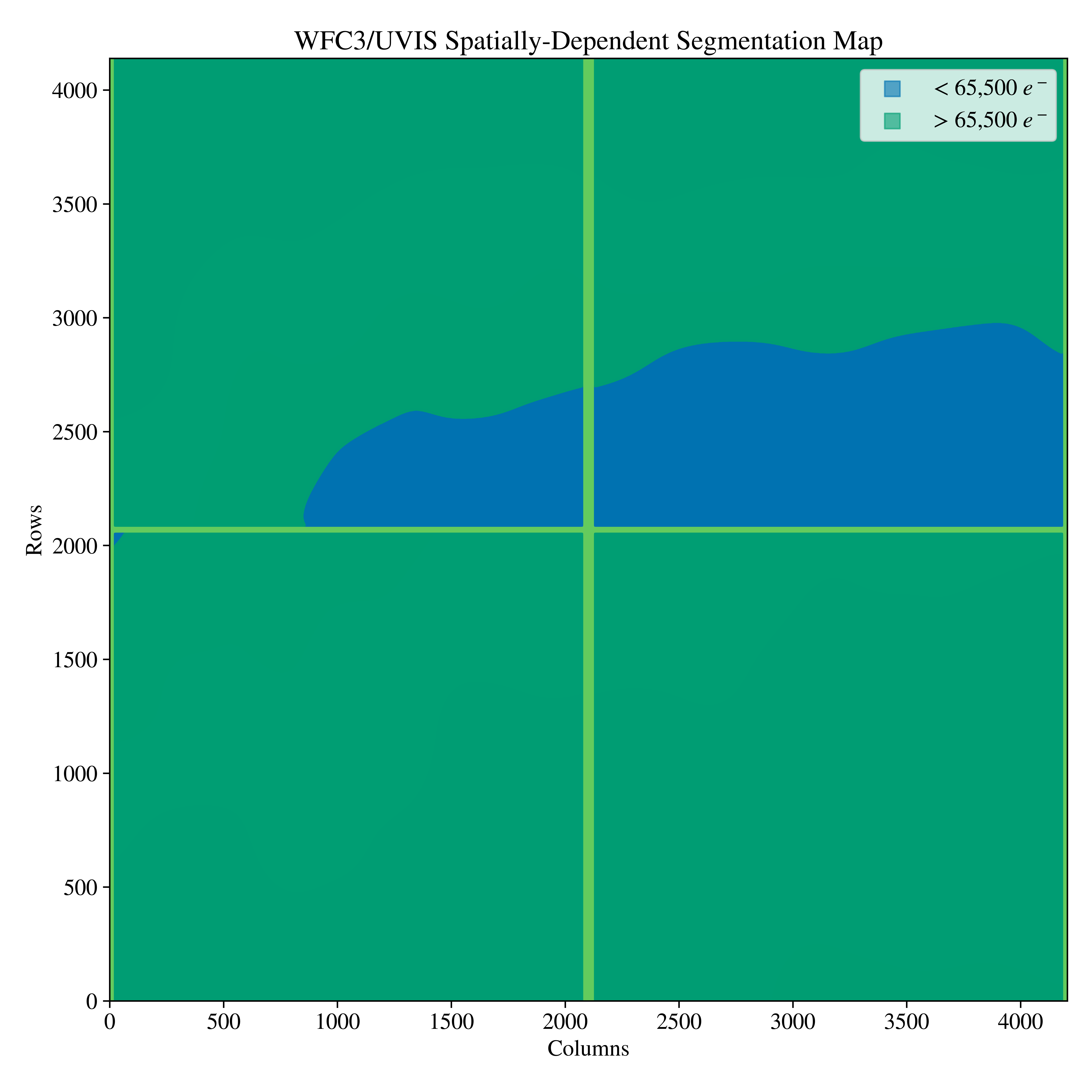}
    \vspace{-2.5em}
    \caption{A segmentation map showing areas where the spatial map thresholds are higher (green) or lower (blue) than the uniform map. Green areas will now have a higher saturation threshold, resulting in fewer saturation flags and regaining science pixels. Blue areas are where the spatial map has a lower threshold, resulting in more pixels being flagged as saturated. Light green areas highlight the overscan regions, which are set to zero. Approximately 87\% of pixels have higher thresholds using the spatial map, while the 13\% with lower thresholds are mostly confined to UVIS1. Almost none of UVIS2 is flagged more by the spatial map ($<$~0.015\% in the upper-left corner). Data reprocessed with the spatial map will have more reliable DQ arrays, recovering usable science pixels over 87\% of the detector.}
    \label{fig:uni_vs_spat}
    \vspace{-0.5em}
\end{figure}

We evaluate the impact on the Data Quality (DQ) flags by processing a subset of UVIS exposures from the WFC3 regression test set using both the uniform and spatial saturation maps. These data are used to validate new \texttt{calwf3} builds and includes a range of exposure times, fields, and filters. These are all full-frame images selected to include numerous bright stars with fluxes near the saturation threshold, in order to exercise the DQ flagging behavior using each map. A summary of the datasets and their parameters is provided in Table~\ref{tab:regression}.

\begin{table}[t!]
    \centering
    \caption{WFC3/UVIS full-frame exposures from the regression test set used to compare the uniform and spatial saturation maps. Datasets were selected to include bright stars near the saturation threshold to test the behavior of DQ flagging with both saturation maps.}
    \label{tab:test_datasets}
    \begin{tabular}{lllll}
        \toprule
        \textbf{Dataset Rootname} & \textbf{Target Name} & \textbf{Filter} & \textbf{Exposure Time (s)} \\
        \midrule
        ib2o01soq                 & NGC-6819              & F606W           & 10                         \\
        ib2o01spq                 & NGC-6819              & F606W           & 600                        \\
        ibc601h2q                 & NGC-6791              & F606W           & 360                        \\
        idhb10esq                 & CL-WESTERLUND-2       & F814W           & 350                        \\
        iem303zaq                 & OMEGACEN              & F502N           & 30                         \\
        \bottomrule
    \end{tabular}
    \label{tab:regression}
\end{table}

\begin{figure}[htb!]
\centering
\centering
\includegraphics[width=\textwidth]{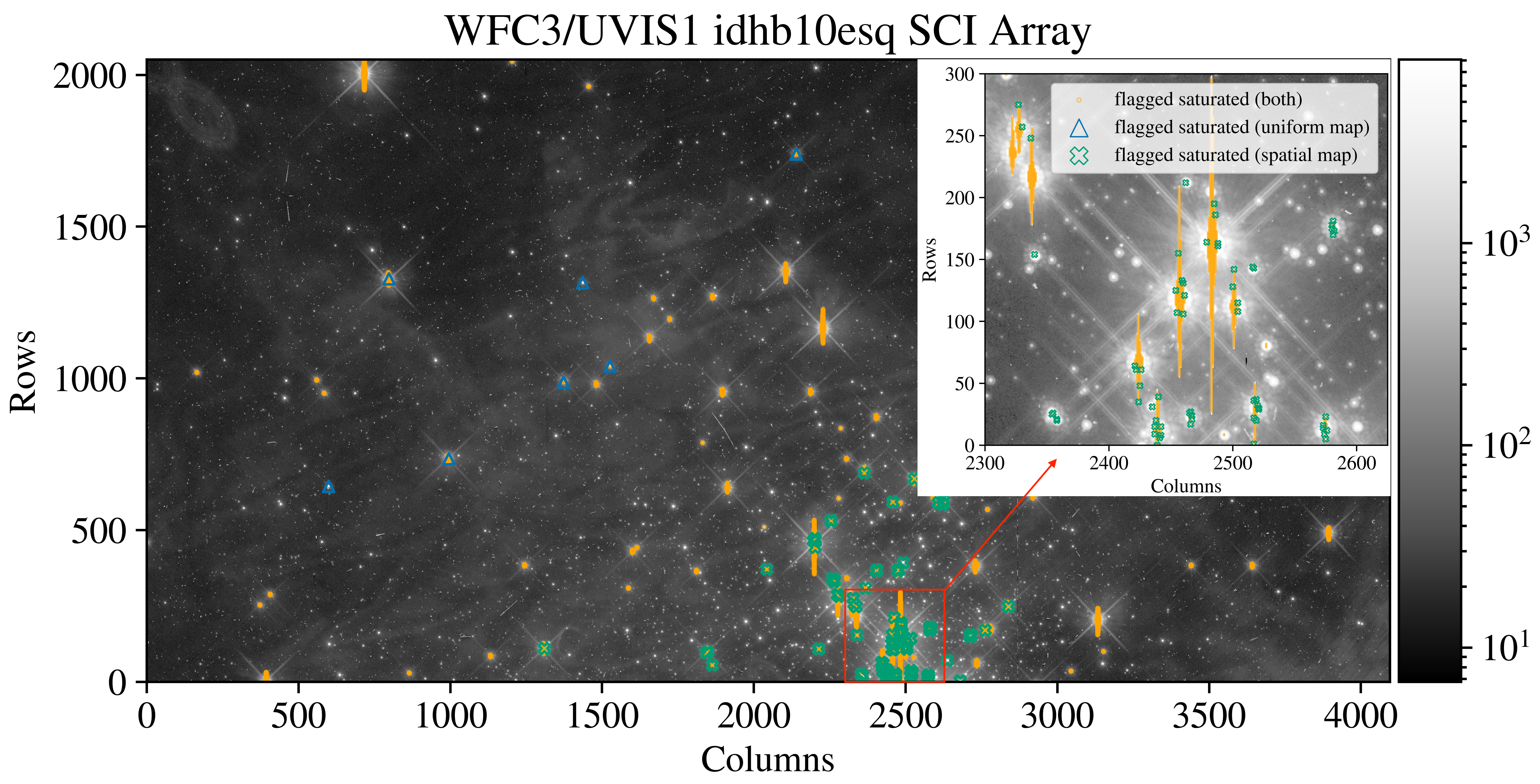}
\includegraphics[width=\textwidth]{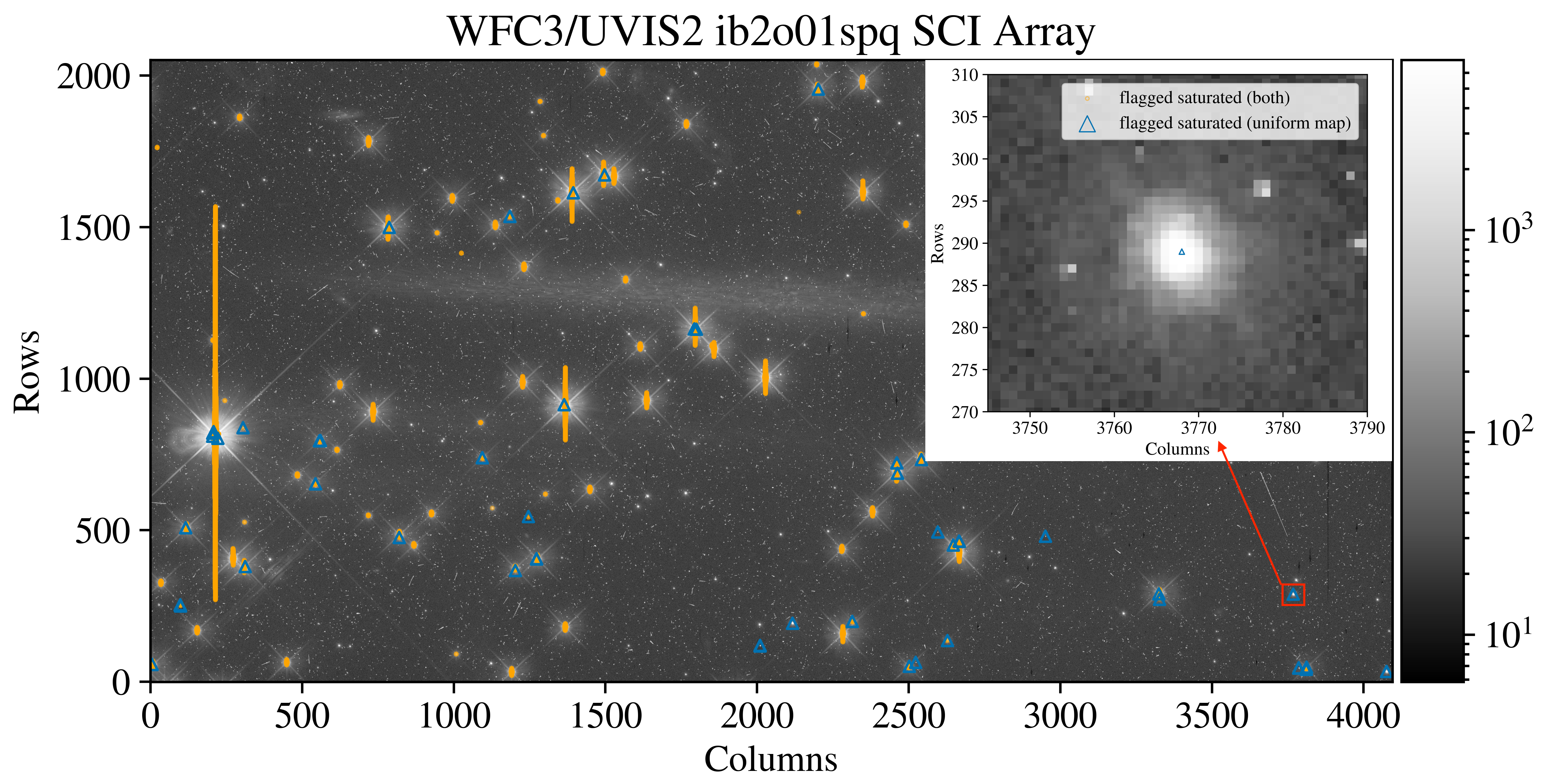}
\caption{A comparison of pixels that are flagged as saturated in the DQ arrays using the uniform map (blue triangles), the spatial map (green crosses), or both (orange circles) overlaid on the science image. The upper panel shows a cluster of stars in UVIS1 within the 13\% region where the spatial map applies lower saturation thresholds than the uniform map. More central pixels are flagged as saturated by the spatial map, particularly along rows affected by charge bleed. The lower panel highlights a bright star in UVIS2, where the spatial map applies higher saturation thresholds than the uniform map. As a result, the core pixel of this star is no longer flagged as saturated using the spatial map. The images are displayed in electrons on a logarithmic stretch with z-scaled limits to emphasize the stellar cores and charge bleed. Across UVIS2 in this example, 64 pixels previously flagged by the uniform map are now unflagged using the spatial map, allowing them to be used in analyses.}
\label{fig:idhb10esq}
\end{figure}

To demonstrate the practical impact of the updated spatial saturation map, Figure~\ref{fig:idhb10esq} presents two examples comparing the flagging behavior of the spatial and uniform maps in \texttt{idhb10esq} and \texttt{ib2o01spq}, respectively. The upper panel of Figure~\ref{fig:idhb10esq} shows a cluster of stars located within the region of UVIS1 where the spatial map applies lower saturation thresholds than the uniform map. In this area, the spatial map flags additional central pixels in the cores of bright stars, consistent with the reduced thresholds. The lower panel of Figure \ref{fig:idhb10esq} highlights a region of UVIS2 containing a bright star. Here, the spatial map applies higher thresholds, resulting in fewer saturated pixels being flagged. For this star, the uniform map would flag it as saturated, whereas the spatial map preserves it. In this case there are no areas on UVIS2 where the spatial map flags more pixels than the uniform map.

Visual inspection of the flagged regions shows that saturation features, such as pixel blooming, are more closely traced when using the spatial map. This behavior helps avoid unnecessary flagging near bright sources to preserve valid data that would otherwise be excluded when combining exposures with the DrizzlePac\footnote{\url{https://www.stsci.edu/scientific-community/software/drizzlepac.html}} software package \parencite{Fruchter2002, Gonzaga2012, Hoffmann2021}. For example, the core of the star shown in the inset of the lower panel of Figure~\ref{fig:idhb10esq} is not flagged with the new saturation map.

\clearpage

\section*{Summary}

In this study, we have developed and implemented a spatially variable saturation map for the WFC3/UVIS detector, replacing the previous uniform threshold approach used in the \texttt{calwf3} pipeline. Leveraging the extensive MAST stellar cutout database and a simple method of detecting PSF shape breakpoints across $\sim$1 million stars observed in the F814W filter, we were able to construct a high-resolution, spatially dependent saturation map. This map captures the full-well depth at the pixel level without relying on paired exposures, aperture photometry, or assumptions about post-saturation linearity. Our analysis divides the detector into 1,024 discrete regions, each 128$\times$128 pixels in size, to robustly fit the saturation breakpoint using piecewise linear models. This procedure enables us to capture both large-scale gradients and fine spatial structure across the UVIS detector.

The resulting saturation map reveals a variation in full-well depth of approximately 13\% across the detector, ranging from 63,465 to 72,356 electrons, consistent with but improving upon the earlier characterization by \textcite{Gilliland2010}. Critically, the updated map shows that 87\% of the detector pixels have higher saturation thresholds than the previous uniform value, allowing the recovery of usable science pixels that would otherwise have been flagged as saturated. We note that reprocessing with this new calibration file will only update the DQ arrays in FLT and FLC products, improving drizzled DRZ and DRC products.

This new, spatially-dependent saturation map reference file has been delivered to CRDS and used to reprocess all WFC3/UVIS data available through MAST. Users are encouraged to either redownload their data or reprocess it with \texttt{calwf3} using the updated saturation map reference file in order to benefit from the more precise pixel flagging. The improved saturation characterization supports more accurate photometric, astrometric, and morphological measurements of bright sources, and contributes to maximizing the scientific return from the rich WFC3/UVIS datasets available in the Archive.

\section*{Acknowledgments}

We thank Ben Kuhn, Joel Green, and members of the WFC3 team for helpful discussions and suggestions that improved the robustness of our analysis and the clarity of this report.

\printbibliography

\clearpage

\appendix

\section*{Appendix: Comparison with \textcite{Gilliland2010}}
In this appendix, we present a direct comparison between the spatially variable saturation map produced by \textcite{Gilliland2010} and the updated map derived in this work. We pursued this comparison to validate our methodology relative to this earlier work, but we explicitly note that exact agreement is not expected. Specifically, our analysis on the FLC level utilizes only the F814W filter, while the sparser datasets available to \textcite{Gilliland2010} required a mix of filters, each with a different flat-field calibration that ultimately raises and lowers some regions of the detector by a few hundred to thousands of electrons. Thus, we focus on the overall variations and global averages in the following comparisons.

\begin{figure}[htb!]
\centering
\vspace{-2em}
\includegraphics[width=0.48\textwidth, trim={1.25em, 1em, 2em, 6em}, clip]{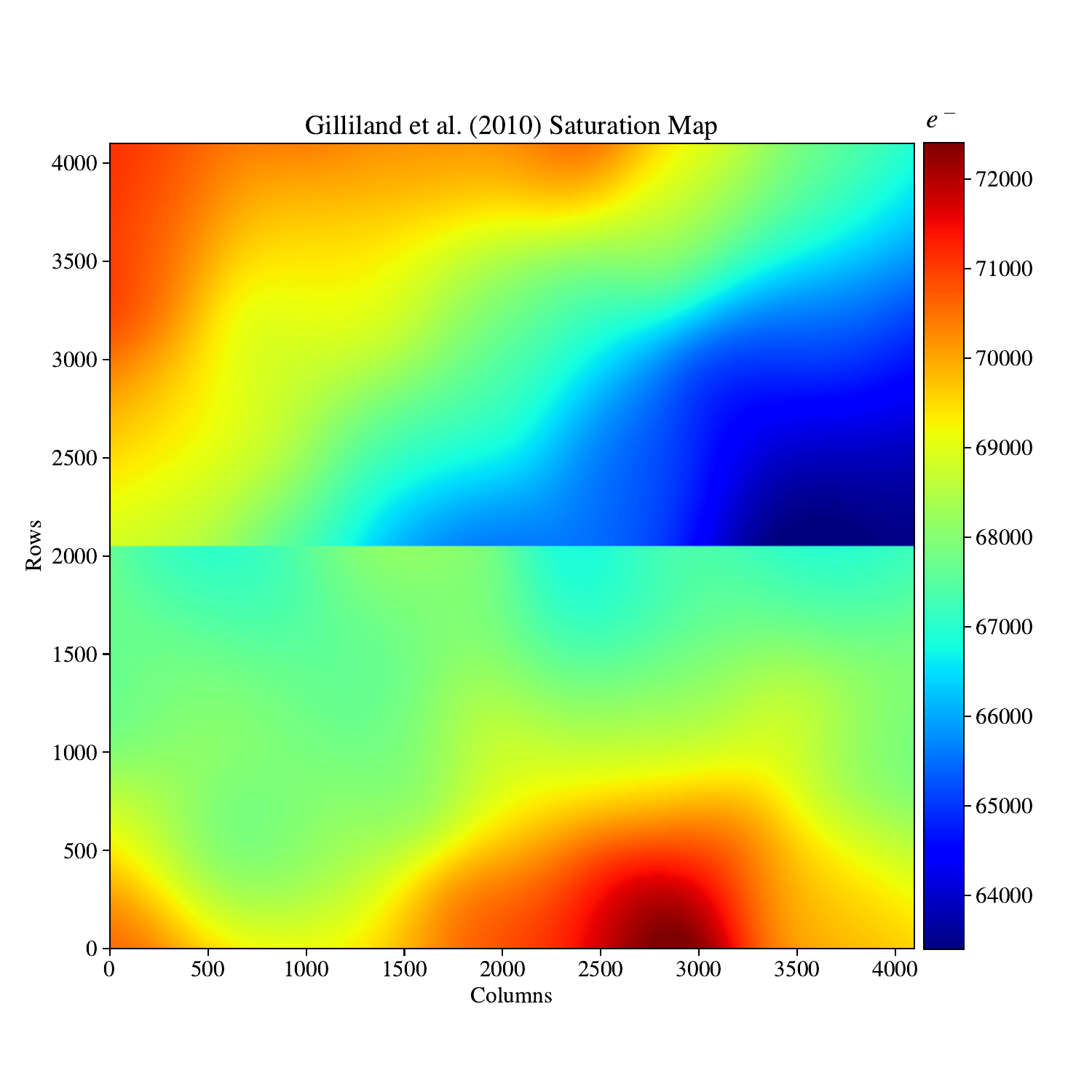}
\includegraphics[width=0.48\textwidth, trim={1.25em, 1em, 2em, 6em}, clip]{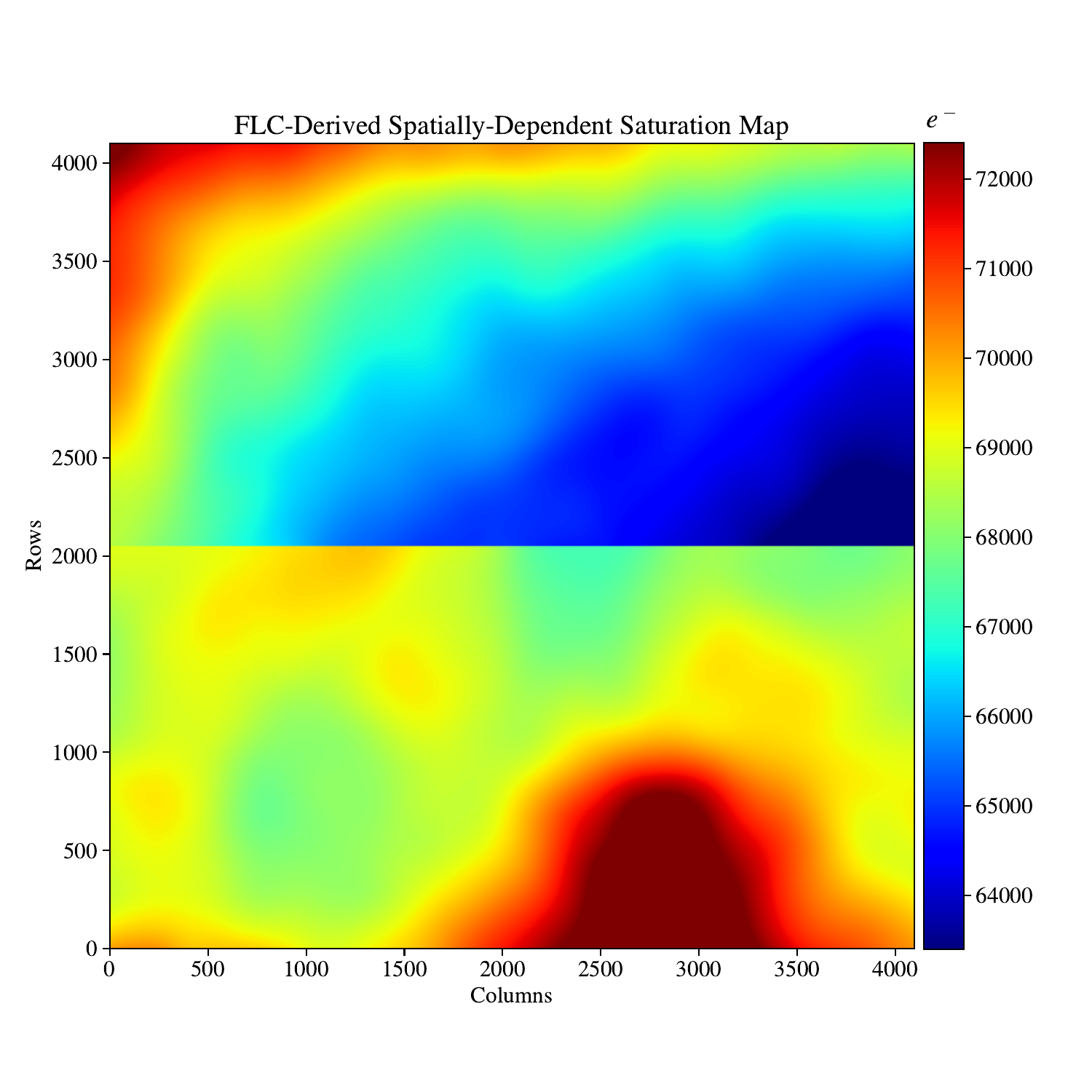}
\includegraphics[width=0.78\textwidth, trim={1.25em, 4em, 2em, 6em}, clip]{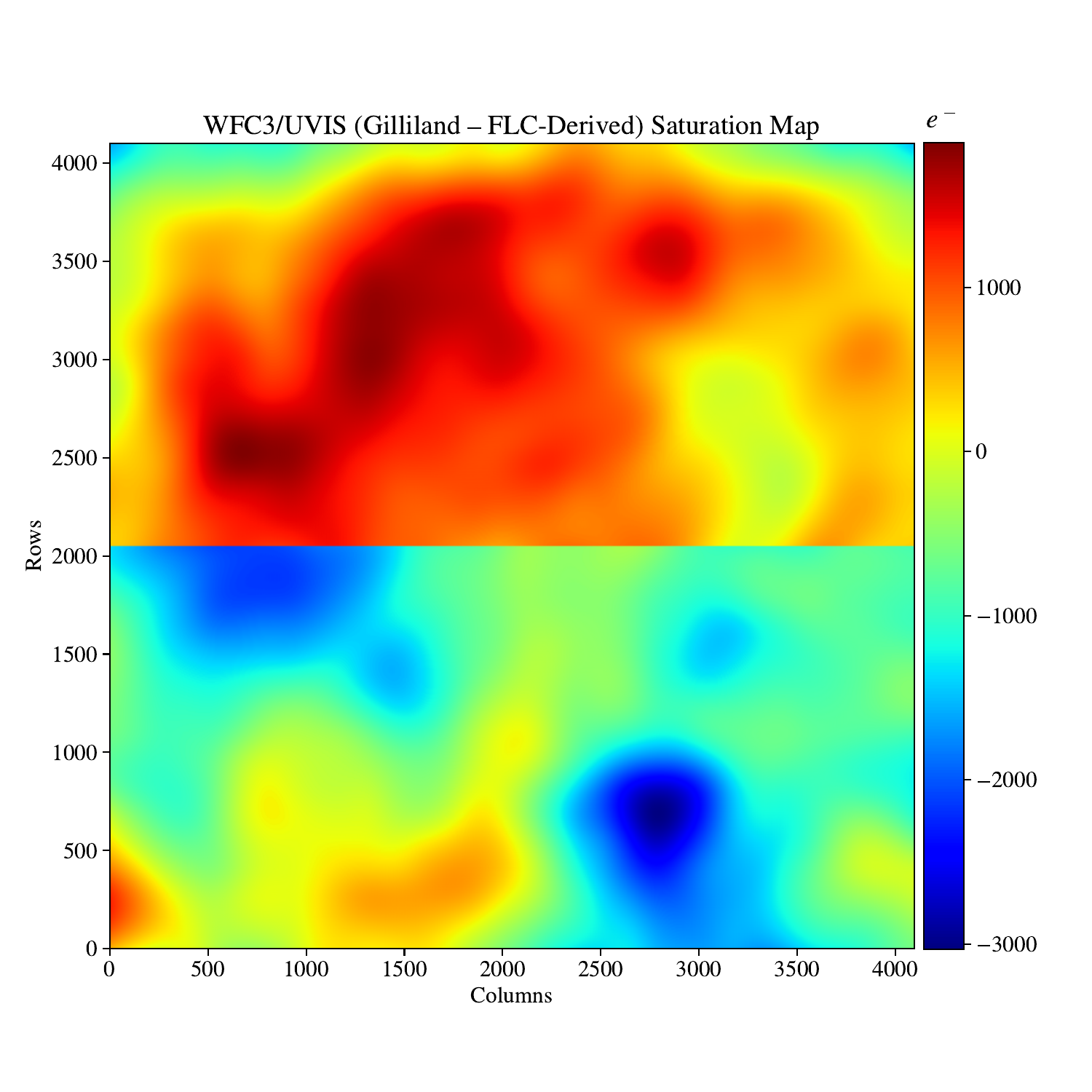}
\vspace{-1em}
\caption{A side-by-side comparison of the saturation map from \textcite{Gilliland2010} (upper-left) and the updated FLC-derived map presented in this work (upper-right) in electrons. Both maps reveal similar large-scale structure in full-well depth across the detector. Bottom: A pixel-wise difference between these maps. Red regions indicate higher saturation values in the older map, while blue regions represent areas where the updated map reports higher full-well depths. The residuals are smooth and large-scale in nature, consistent with physical structure across the CCDs. Notably, UVIS1 shows overall positive differences, whereas UVIS2 shows mostly negative differences. An exact match is not expected due to differences in the filters and calibrations utilized in the analyses, as discussed in the main text.}
\label{fig:gill_vs_spat}
\end{figure}

\begin{figure}[htb!]
\centering
\includegraphics[width=\textwidth]{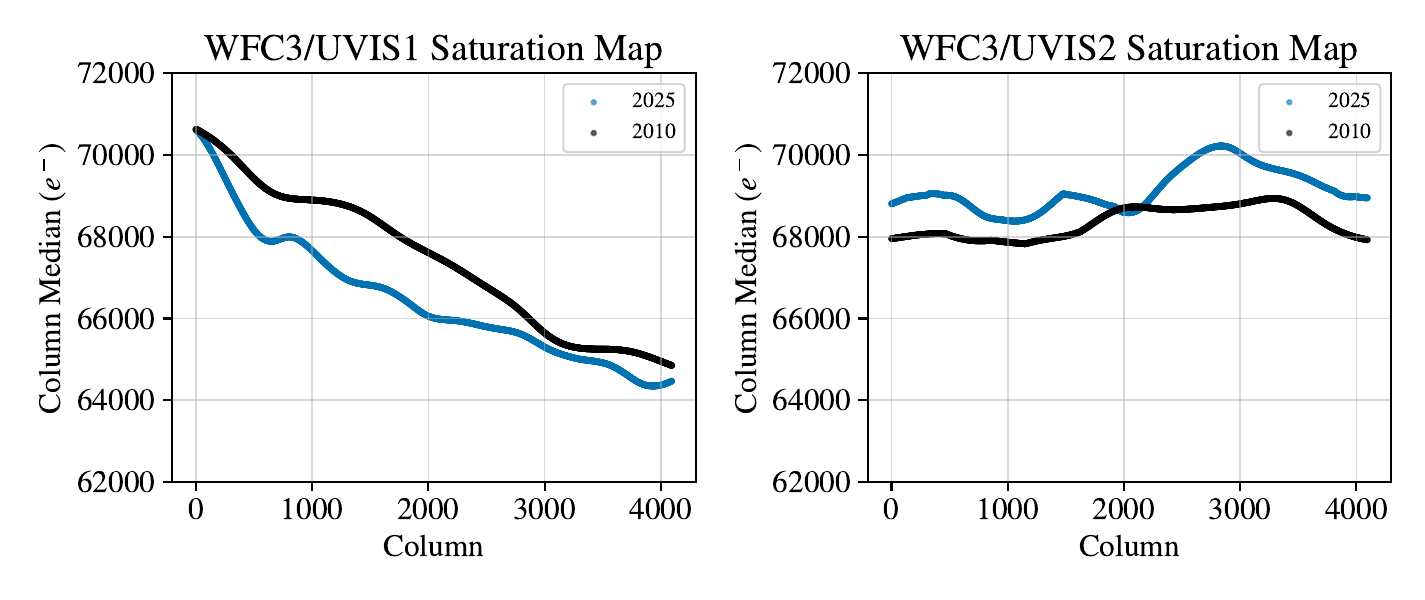}
\vspace{1em}
\includegraphics[width=\textwidth]{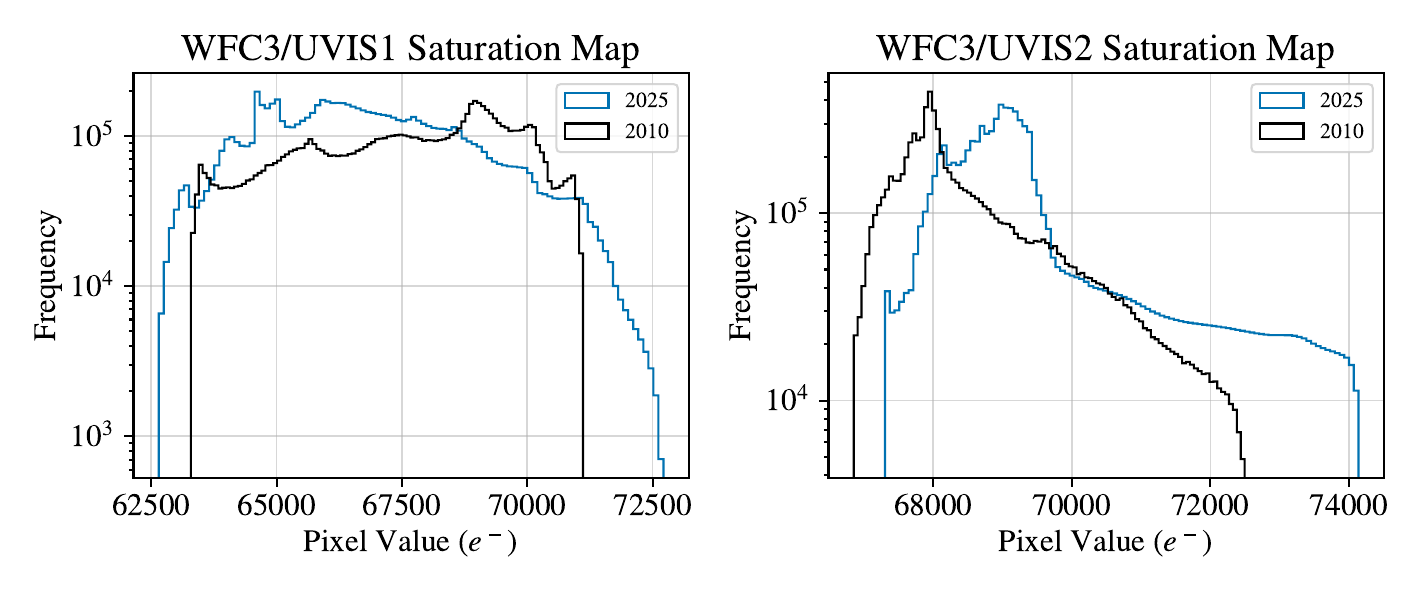}
\caption{Top: Column-wise median saturation values computed from the Gilliland et al. (2010) map (black) and the updated spatial map presented in this work (blue, labeled 2025), shown separately for UVIS1 (left) and UVIS2 (right). Both maps display a consistent large-scale gradient in saturation level across the detector, likely reflecting inherent structural differences in the CCD. The updated map also resolves finer-scale structure, made possible by the larger number of stars and finer spatial sampling used in its construction.
Bottom: Histograms of the saturation values from both studies for UVIS1 and UVIS2. In UVIS1, the distributions are broadly similar, though the Gilliland map shows sharp cutoffs at both the minimum and maximum values. In UVIS2, both maps exhibit a high-value tail corresponding to the ``happy bunny" region, with the spatial map shifted toward higher saturation levels.}
\label{fig:column_medians_and_histograms}
\end{figure}

First, we generated our saturation map using calibrated, CTE-corrected (FLC) images, enabling a direct comparison to the Gilliland map, which was constructed using uncorrected (FLT) images with data from early in the mission that suffered minimal CTE losses. As shown in the top panels of Figure \ref{fig:gill_vs_spat}, both maps exhibit similar large-scale structure across the detector, with variation on the order of 10\%. The bottom panel shows the pixel-by-pixel difference between the two maps, which reveals localized differences that are generally small, with most regions differing by less than $\sim$1000 electrons.

One key calibration difference between the analyses should be emphasized. Prior to mid-2016, UVIS2 fluxes were not normalized to match UVIS1—each chip had independent zeropoints. With the introduction of the FLUXCORR step in \texttt{calwf3} v3.3, UVIS2 images are scaled to match UVIS1 using the \texttt{PHTRATIO} keyword, defined as the ratio of UVIS2 to UVIS1 inverse sensitivities (\texttt{PHTFLAM2/PHTFLAM1}). See the Data Handbook, Chapter 9.1, Photometry \parencite{Pagul2024} for more information. This scaling is applied in producing FLT/FLC images and therefore directly affects the FLC-level PSF cutouts used in this analysis. In practice, the stored pixel values for UVIS2 are multiplied by \texttt{PHTRATIO} relative to their 2010 calibration, likely contributing to some of the observed differences between the 2010 and 2025 saturation maps, particularly for UVIS2. This systematic offset is measurable, varying with time and filter \parencite{Calamida2021}.

In the present 2025 analysis, cutouts are drawn exclusively from the F814W filter but span all available epochs, so the time-dependent \texttt{PHTRATIO} scaling is applied. For F814W, a representative \texttt{PHTRATIO} at MJD=55008 (2009-06-26) is 1.017 (\textcite{Calamida2021}, Table 8), which when applied to a 63,000~e$^{-}$ saturation level corresponds to a shift of 1,071~e$^{-}$. Because sensitivity changes differ slightly between UVIS1 and UVIS2, the count-rate ratio may vary by up to 2\%, implying that UVIS2 saturation levels can differ by roughly 1,000–2,000~e$^{-}$ across epochs purely from \texttt{PHTRATIO} scaling. The application of the \texttt{PHTRATIO} scaling to UVIS2 in our analysis, absent in the Gilliland et al. (2010) analysis, likely accounts for the larger deviations seen for UVIS2 in the difference map in Figure~\ref{fig:gill_vs_spat}. This effect should be considered when comparing the two chips or when working across epochs and filters.

To better understand the spatial characteristics of these differences, we compute column-wise medians of both maps and plot them across the detector (top of Figure \ref{fig:column_medians_and_histograms}). These plots confirm that the large-scale gradient in saturation level, likely due to inherent structural differences across the CCD, is captured consistently by both methods. However, the new fitting procedure allows us to resolve finer-scale variations in full-well depth, thanks to the larger volume and broader spatial coverage of the data used in this work.

The bottom panel of Figure~\ref{fig:column_medians_and_histograms} shows histograms of saturation values for UVIS1 and UVIS2. In UVIS1, the distributions from both maps are broadly similar, although the Gilliland map exhibits sharp cutoffs at the minimum and maximum values—an effect not typically seen in Gaussian smoothed distributions. This suggests that the Gilliland map may have imposed boundaries or excluded edge cases in its construction. In UVIS2, both maps show a high-value tail, with the spatial map shifted slightly toward higher saturation values. This excess corresponds to an isolated region of elevated values—commonly referred to as the “happy bunny” feature—in the UVIS2 spatial map and reflects a thin region of the detector as shown in Chapter 5.6, Figure 5.19 of the WFC3 Data Handbook \parencite{Pagul2024}. This area appears as a distinct structure in the WFC3/UVIS flat-field image and corresponds to increased full-well capacity and higher saturation thresholds.

This broad consistency between the maps lends confidence to the methodology used in the current study, while alleviating any dependence on aperture definition, offsets in spatial position or sky level between exposures, linearity beyond saturation, and the effects of cosmic rays thanks to outlier rejection during fitting, which were inherent to the earlier analysis. We note that the \textcite{Gilliland2010} analysis produced a high-quality map utilizing the sparse datasets available at that time, and the current analysis builds upon that excellent work to characterize the WFC3/UVIS saturation on smaller scales, using a simple and robust methodology.

\end{document}